\documentstyle[aas2pp4]{article}

\def\Msun{\ifmmode M_{\odot} \else $M_{\odot}$\fi}
\def\Lsun{\ifmmode L_{\odot} \else $L_{\odot}$\fi}
\def\eg{{\it e.g.,\ }}
\def\ie{{\it i.e.,\ }}
\def\etal{{et al.~}}

\lefthead{Chatzichristou}
\righthead{IR-Warm Seyfert Galaxies}

\begin{document}

\title{Multicolour Optical Imaging of IR-Warm Seyfert Galaxies.\\
    II. Optical and IR Properties and their Correlations}

\author{Eleni T. Chatzichristou}
\affil{Leiden Observatory, P.O. Box 9513, 2300 RA Leiden, The Netherlands}

\affil{NASA/Goddard Space Flight Center, Code 681, Greenbelt, MD 20771}



\begin{abstract}
This paper is the second in a series, studying the optical properties of a
sample of mid-IR Warm Seyfert galaxies and a control sample of mid-IR Cold 
galaxies. The present paper is devoted to aperture photometry.
We discuss nuclear (within the central 2 kpc) and disk optical properties 
and their correlations with IR properties. We find a transition in 
the observed optical and IR properties from the Cold to the Warm Seyfert 2 and
Seyfert 1 samples, with a partial overlap between the first two.
This is in the sense of: (i) increasing
nuclear optical luminosities and bluer nuclear colours (ii) 
decreasing disk optical luminosities and sizes
(iii) increasing 25 $\mu$m luminosities, decreasing far-IR excess 
(\(\frac{L_{FIR}}{L_B}\)) and IR-loudness  ($\alpha_{(V,25)}$,
$\alpha_{(V,60)}$) and bluer $\alpha_{(25,60)}$ and $\alpha_{(12,100)}$ 
colours. We interpret these results as indicating larger dust content, and
disk star formation for Seyfert 2s, while the optical properties of Seyfert
1s are mostly dominated by their nuclei. The 25 $\mu$m emission of the Warm 
sample is mainly due to the AGN thermal component, although in Seyfert 2s it
is probably
further enhanced by disk star formation. Their far-IR (60-100 $\mu$m) luminosity
on the other hand, traces mainly emission at large scales (host disks).
In the Cold galaxies, the bulk of IR emission at all wavelengths appears to be
dominated by dust in their disks.

\end{abstract}




%

\section{Introduction and Data Preparation}

In \cite{paper1} (hereafter Paper I) we have presented our sample of IR-Warm 
Seyferts, selected from the De Grijp \etal original sample of warm IRAS 
sources (\cite{grijp87} and \cite{grijp92}) and a control sample of 
IR-Cold galaxies, that span similar redshift and luminosity ranges as the Warm
sample. The basic data, $B,V,R,I$ broad-band optical imaging and reduction 
procedures were summarized in Paper I. In the present Paper II we extract 
global photometric measures for the sample galaxies and compare the Seyfert 1 
vs Seyfert 2 and the Warm vs Cold (sub)samples.

Among the 54 Seyfert galaxies in our Warm sample photometric measures are 
available for 33 objects and the results are presented in Table~\ref{tab1}.
Among the  16 galaxies in our Cold sample, we have photometric data for 14 of 
them, presented in Table~\ref{tab2}.
For the rest of the objects the observing conditions were too variable
to derive accurate colours and thus are omitted from further discussion in the
present paper. Through-out this paper (as in Paper I) the objects are 
identified with both their IRAS name and their serial number (with prefix I) in 
the original list (\cite{grijp87}).

The present paper is structured as follows: in Section 1 we
summarize the data preparation and in particular the identification of the
optical counterparts of the IRAS sources in the case of multiple systems.
In Section 2 we describe the results of our aperture photometry, 
concentrating in intercomparisons between the (i) Warm Seyfert 1 and 2 
subsamples and (ii) the Warm and Cold (sub)samples; we then summarize the main
conclusions of that section. In Section 3 we do an exhaustive search of
possible relations between optical and IR properties and intercompare the
Warm and Cold (sub)samples. We summarize our main conclusions at the end
of that section. We close in Section 4 with some concluding remarks in an 
attempt to understand the nature of IR-Warm seyferts based on our aperture 
photometry results.

\subsection{Optical Identifications}

Inspection of the broad-band data presented in Paper I shows a large number of
closely interacting or double nucleus systems. It is important to identify 
which member of each system is the optical counterpart of the IRAS source.
In the original list of warm sources (\cite{grijp87}), the optical counterparts 
were identified  using the Palomar Sky Survey and the ESO/SRC plates. The 
limited angular resolution on those plates often prevented detection of the 
duplicity of the sources. In fact, in some cases where the two members of an 
interacting system were resolved, the IRAS position lay between the two 
galaxies. The IRAS angular resolution varies between 0.5 arcmin (at 12 $\mu$m) 
to 2 arcmin (at 100 $\mu$m) and the positional accuracy depends on the source 
size, brightness and spectral energy distribution.
To resolve the ambiguity for our sources, we have re-calculated the IRAS 
positional accuracy for all fields from the individual IR images (flux and 
position calibrated) extracted from the SRON IRAS database (Groningen, The 
Netherlands). Typically, accuracies are of the order $\sim$10-30 arcsec
along each direction. For the (twelve) ambiguous cases, with more than one 
optical candidates for the IRAS identification, optical and IR positions were 
then compared. Details are given in \cite{thesis}. After this analysis, there
were three remaining cases with ambiguous identification:
the Warm IRAS 11298+5313 and the Cold IRAS 04530-3850 and IRAS 09406+1018
galaxies.
For each of these systems, both members are retained as possible IRAS candidates
and are included in subsequent analysis of our data.
 
\subsection{Double Nucleus Systems}

In both samples there is a number of galaxies where more than one nuclei 
are visible, embedded in a common envelope. Among the objects with photometric 
data, we have three such cases in the Warm sample: two Seyfert 2s, 
IRAS 13536+1836 (I333) and IRAS 19254-7245 (I489), one Seyfert 1, IRAS 
19580-1818 (I495) and one Cold system, IRAS 07514+5327 (I231). IRAS 13536+1836 
has been extensively studied in the past, its eastern Seyfert nucleus being a 
radio source and most probably the main source of IR emission (see \eg 
\cite{eleni95}). The nature of the western nucleus is less clear, probably a 
lower ionization Seyfert or a LINER (previous reference). Consequently, 
although in our tables we are listing measures for both nuclei, in our 
histograms and plots we are only considering the east nucleus. The case of IRAS 
07514+5327, is more ambiguous. \cite{eleni98} showed that both condensations 
are extended and show \ion{H}{2}-like spectra. In the above paper, arguments are given
for the southern condensation to be the locus of the (highly obscured) galaxy 
nucleus. However, it is possible that the northern condensation is the location
of a second nucleus and thus, both are included in our analysis. There is one 
more ambiguous case among the Cold sample objects, IRAS 04265-4801, with a 
peculiar patchy morphology and several knots of emission in the central region.
The brightest central knot is the dominant (if not unique) nucleus and thus
the one used in our subsequent analysis.
In the remaining two cases of Warm double nucleus objects listed above, we have
used measures for both nuclei. In the histograms we adopt a mean value
for the nuclear magnitudes and colours and in the rest of the plots we show
data points for both nuclei, connected with a line. We use 
total magnitudes and colours for the system as a {\em whole}, since
in none of the above cases we can distinguish two individual
stellar systems. It is interesting that in these cases, the total magnitudes 
of the merger system are not significantly brighter that the mean total magnitudes for each sample 
and sometimes are actually fainter. Because of the relatively small number of
double nucleus systems in our samples, the above assumptions are unlike to 
affect our results from the statistical correlations.

\section{Aperture Photometry Results}

To obtain magnitudes and colours for our galaxies, we have used both 
(i) concentric circular apertures at specified radii and (ii) elliptical 
apertures with varying centers that were fitted to the galaxy's isophotes at 
increasing radii (the photometric algorithm used is described in \cite{thesis}).
The two methods give sometimes significantly different results; detailed comparisons and
discussion can be found in \cite{thesis}.
A number of photometric corrections to the measured magnitudes were applied, 
that are described in detail, also in \cite{thesis}.

In Tables~\ref{tab1} and ~\ref{tab2} we list magnitudes and colours for our 
sample
objects with photometric information. These are referred to as ``total'' when
measured within the $\mu_{B}$=25 mag/arcsec$^{2}$ elliptical aperture 
and as ``nuclear'' when measured within an elliptical aperture of 
semi-major axis 2 kpc. The choice of the nuclear radius is somewhat arbitrary; 
large enough so as to not be affected by seeing effects and at the same time 
sufficiently small so that properties within this region should be dominated 
by the central nucleus.
Magnitudes, colours and major axis scale lengths are corrected for 
Galactic extinction and relativistic effects as described in \cite{thesis}.
The objects in Tables~\ref{tab1} and ~\ref{tab2} are listed in order of 
decreasing mid-IR warmness $\alpha_{(25,60)}$. For each object, the first row
represents total and the second nuclear magnitudes.

\begin{table}
\dummytable\label{tab1}
\end{table}

\placetable{tab2}

\subsection{ Seyfert 1 vs Seyfert 2 Galaxies }

\begin{figure}
\epsscale{1.5}
\plotfiddle{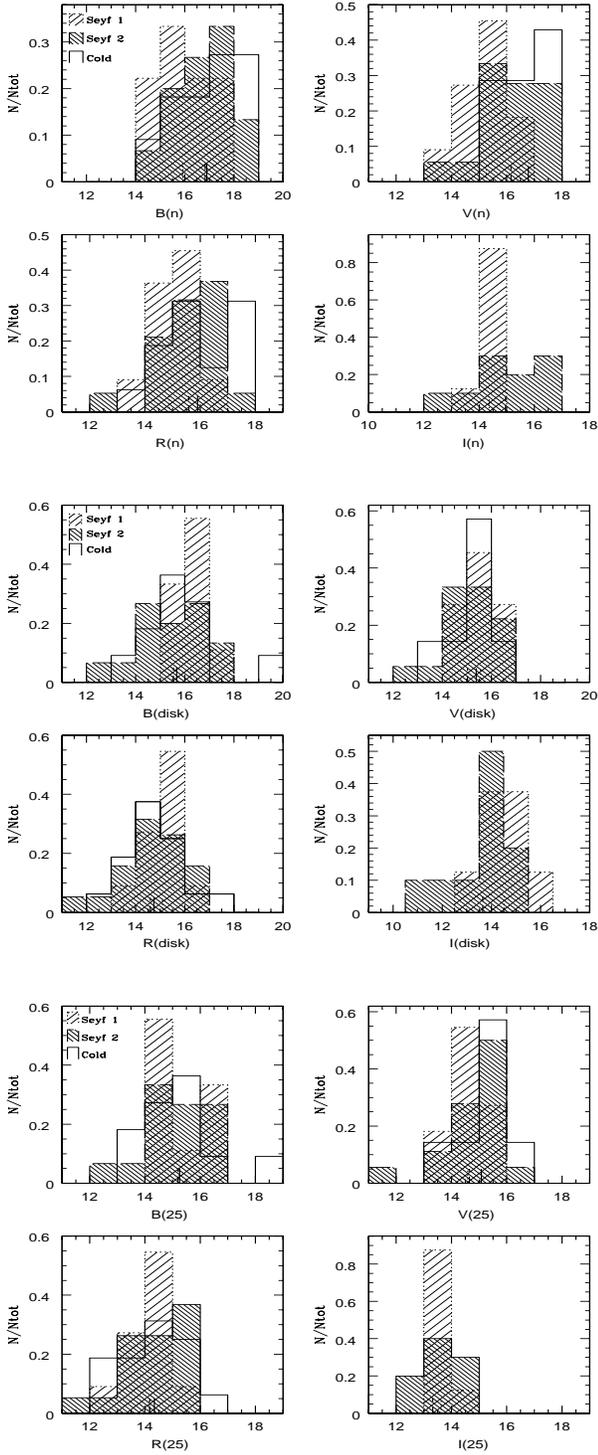}{400pt}{0}{85}{90}{-260}{-90}
\caption{Fractional distributions of nuclear, disk and total apparent magnitudes, for the Warm Seyfert 1 and 2 and the Cold subsamples. Mean values are indicated as dotted, dashed and solid vertical bars, respectively, on the lower x-axes. \label{f1}}
\end{figure}

In Figures~\ref{f1} and~\ref{f2} we plot the fractional 
distributions of nuclear, disk and total (apparent) magnitudes and colours for 
the Warm Seyfert 1 and 2 subsamples and for the Cold sample. The subscript 
{\em disk} denotes colours and magnitudes within the annulus defined between 
semi-major axes 2 kpc - a$_{25}$.
Filled histograms represent the Warm Seyfert 1 and 2 distributions and empty 
histograms the Cold galaxy distributions. The sample mean values are shown as 
vertical bars on the lower x-axes, as explained in the figure captions.
The Table~\ref{tab3} lists median and mean values for all the measured 
quantities and the various subsamples. 

\placefigure{f1}

Here are our main conclusions:

\subsection*{\it Magnitudes}

(i) The nuclear magnitudes show a significant shift in their distributions: 
Seyfert 1 nuclei are brighter than their Seyfert 2 counterparts and there is
a tendency for this discrepancy to increase as we progress to shorter 
wavelengths. The Student's T-test shows statistically different means for the
two subsamples, while the K-S test shows that the hypothesis that the two 
distributions match remains less than 10\% (not a statistically significant
level) for all $B,V,R,I$ magnitudes. The above result persists for both 
apparent
and absolute magnitudes, confirming that there are no distance-selection
effects for Seyfert types 1 and 2 in our sample. A straightforward 
interpretation of this result is, according to the orientation 
unification model, that in Seyfert 1s the AGN is visible while in Seyfert 2s 
it remains (partially) obscured by dust. Another possibility is however, that 
the Seyfert 1 nuclei are intrinsically brighter and bluer compared to the 
Seyfert 2 nuclei.

(ii) Although the distributions of disk luminosities are comparable between the
two Seyfert subsamples there is consistently a tendency for the Seyfert 2 
disks to be brighter, although not at a statistically significant level.
If this tendency is however true, it could indicate larger star formation
in the Seyfert 2 disks. Another possibility is that the disk luminosity simply
scales with the larger size of the Seyfert 2 host galaxies, which will be 
discussed below. None of these alternatives is however compatible with the 
orientation unification scheme for Seyferts.

(iii) The Warm Seyfert 1 and 2 galaxies have comparable {\em total} 
optical fluxes, the two opposite effects described in (i) and (ii) compensating
for each other and this seems to be independent of (optical) wavelength. The 
F-Variance and the T-Student's test show indeed statistically similar variances
and means for the two samples.

(iv) In Figure~\ref{f3} we show mag-mag plots for our three samples.
The trends described earlier are better visible here:

{\em Nuclear vs Disk}: Seyfert 2 nuclei tend to be fainter and to 
reside in brighter disks, than their Seyfert 1 counterparts. Also, for both 
samples nuclear and disk colours correlate (\ie brighter nuclei tend to reside
in brighter disks) this relation being better defined and steeper for Seyfert 
2s. 

{\em Nuclear vs Total}: The tendency for the nuclear magnitudes to scale with 
total magnitudes is well defined in these diagrams, in particular for the 
Seyfert 1s indicating that their total optical fluxes are dominated by their 
nuclei.

\subsection*{\it Colours}

\begin{figure}
\epsscale{1.5}
\plotfiddle{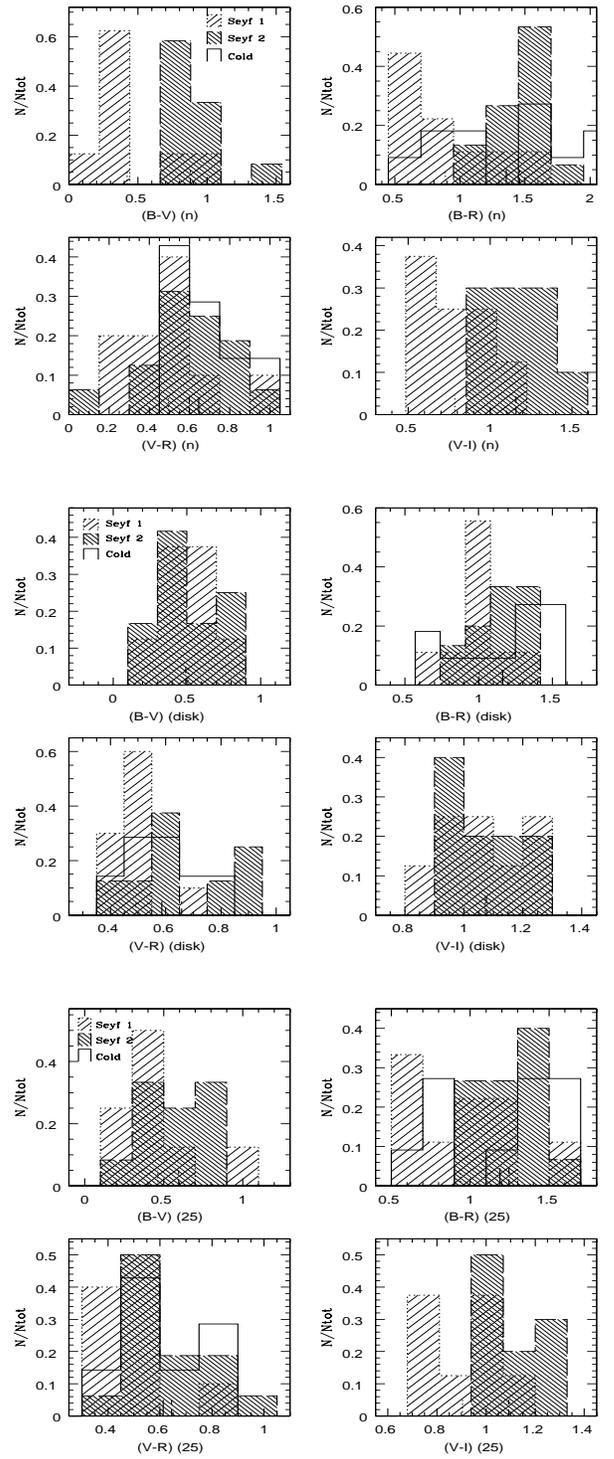}{400pt}{0}{85}{90}{-260}{-90}
\caption{Fractional distributions of nuclear, disk and total colours, for the Warm Seyfert 1 and 2 and the Cold subsamples. Mean values are indicated as dotted, dashed and solid vertical bars, respectively, on the lower x-axes. \label{f2}}
\end{figure}

\placefigure{f2}

(v) The distributions of colours shown in Figure~\ref{f2} indicate
clear differences between the two Seyfert (sub)samples. The Seyfert 1s 
have bluer nuclear colours compared to the Seyfert 2 nuclei. A similar shift
is seen for their total colours. 
For the disk colours any such shift is less pronounced if at all
present, this indicating that the Seyfert total colours are mainly reflecting 
the colours within 2 kpc from the center, this being true in particular for the Seyfert 1s, as we 
shall show below. Moreover, these results indicate that the two Seyfert types
must have dramatically different colour gradients. These will be indeed 
extensively discussed in Paper IV.
The Student's T-test shows in general significantly different means for the 
distributions of nuclear and total colours for the two (sub)samples.
Furthermore, the K-S test lends a 95\% statistical significance to the 
hypothesis that the Seyfert 1 and 2 nuclear and total colour distributions are 
not drawn from the same parent population. 
The exception to this is a large overlap of the $(V-R)$ nuclear colours, which 
however can be explained by the contamination of the Seyfert 1 $R$-band 
magnitudes by nuclear broad $H_{\alpha}$ emission, preferentially reddening 
their $(V-R)$ colours. Another possible effect is contamination of the $V$-band
magnitudes by the [\ion{O}{3}]$_{4959,5007}$ emission lines, that are more likely
to affect the (shorter baseline) $(V-R)$ colours. This effect would lead to an 
artificially bluening of both the Seyfert 1 and Seyfert 2 colours by similar 
amounts, within the orientation unification scheme.

\placefigure{f3}

\begin{figure}
\epsscale{1.0}
\plotone{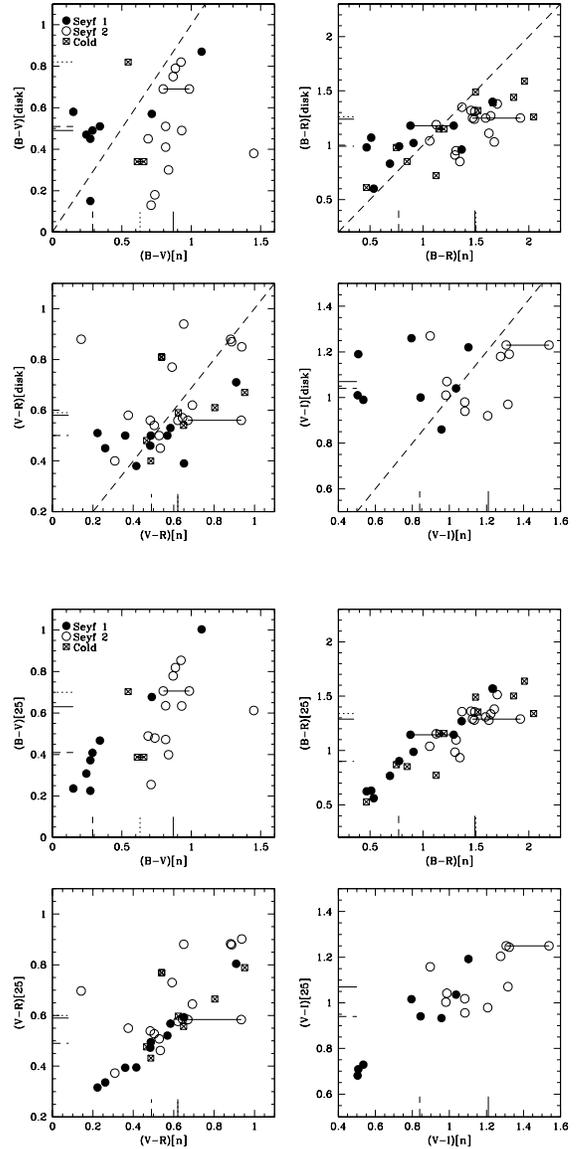}
\caption{Nuclear vs Disk and Nuclear vs Total colour-colour plots for the Warm Seyfert 1 and 2 type galaxies and for the Cold sample. Horizontal lines connect measures for the two nuclei of double nucleus systems. Labels 1 and 2 indicate Seyfert type 1 and 2 nuclei within the Cold sample galaxies. Bars indicate median values for the three samples: dashed for Seyfert 1s, solid for Seyfert 2s and dotted for the Cold galaxies. Dashed lines indicate the loci of equal nuclear and disk colours. \label{f4}}
\end{figure}

(vi) In Figure~\ref{f4} we show nuclear vs disk and nuclear vs total 
colour-colour plots. There seems to exist a correlation between the nuclear and
disk colours for the Warm Seyfert 2 galaxies (with the exception of the $(V-I)$
colours; $(B-R)$ correlation coefficient 0.47, at 0.07 significance level).
A similar trend is seen for the $(B-R)$ colours in Seyfert 1s. This  
basically indicates that redder Seyfert 2 nuclei reside in redder hosts, the 
most straightforward interpretation being that reddening due to dust affects 
similarly their nuclear and outer regions.
In the nuclear vs total colour plots we find significant correlations for both 
samples that become very tight in the case of Seyfert 1s: the correlation
coefficients (Spearman's test) for the $(B-R)$ and $(V-R)$ colour-colour plots 
are $\sim$1 and $\sim$0.7 for the Seyfert 1s and 2s, respectively, at 
significance levels better than 0.003.
Thus, we find again that the Warm Seyfert total galaxy colours are dominated by
their colours within 2 kpc from the nucleus, in particular in Seyfert 1s.  
 
The most important effect depicted in Figure~\ref{f4} is that, while 
the disk colours of both types 1 and 2 objects are mostly overlapping, the 
Seyfert 2 nuclei are significantly redder (by 0.1-0.6 mag) than the Seyfert 1 
nuclei. In other words, the Seyfert 2s tend to have redder nuclei for similar 
disk colours. This is indeed consistent with the nuclear torus obscuration 
hypothesis for Seyferts. However, we have also shown that Seyfert 2 disks tend
to be somewhat redder than their Seyfert 1 counterparts (by an average $<$0.2
mag) and that the Seyfert 2 disk and nuclear colours tend to correlate 
(Figures~\ref{f2} and ~\ref{f4}) with each other. These indicate 
that Seyfert 2s are systems with more overall reddening 
(larger amount of dust) than Seyfert 1s.

(vii) The dashed line in Figure~\ref{f4} indicates the loci of similar 
nuclear and disk colours. Points lying to the right/left hand side of it 
indicate respectively redder/bluer nuclei compared to their surrounding disks. 
With the exception of the $(V-R)$ colours (possible reasons for that are 
discussed above) all the other colour-colour diagrams show positive colour 
gradients (\ie redder colours as a function of radius) for most Seyfert 1s and 
{\em opposite}, negative colour gradients for most Seyfert 2s. Colour gradients
are the subject of Paper IV, where they will be discussed in detail.

\placefigure{f4}

\placetable{tab3}

\subsection*{\it Host Sizes}

\begin{figure}
\epsscale{0.5}
\plotone{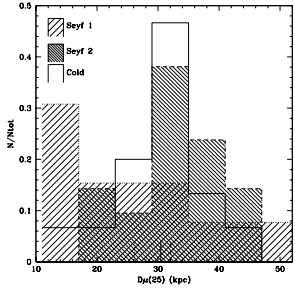}
\caption{Fractional distributions of major axis diameters, measured within the isophote corresponding to $\mu_{B}$=25 mag arcsec$^{-2}$, for the Warm Seyfert 1 and 2 and the Cold samples. Mean values are indicated by the dotted, short and long solid vertical bars, respectively, on the lower x-axis. \label{f5}}
\end{figure}

\begin{figure}
\epsscale{1.0}
\plotone{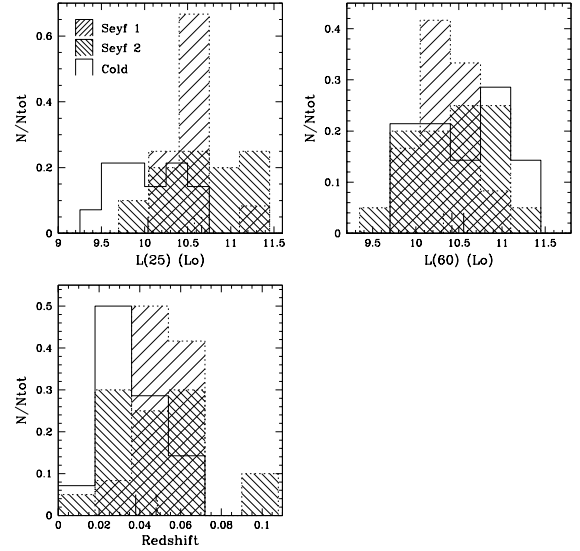}
\caption{Fractional distributions of the 25 and 60 $\mu$m flux densities and the redshifts, for the Warm type 1 and 2 Seyferts and the Cold sample (only the objects with photometric information). The vertical bars on the lower x-axes indicate mean values (dotted for Seyfert 1s, dashed for Seyfert 2s and solid for Cold galaxies). \label{f7}}
\end{figure}

(viii) In Figure~\ref{f5} we show the distribution of major-axis
diameters corresponding to the $\mu_{25}$ isophotes, mostly measured on the
$B$-band images and, when these were not available, on the $V$-band images.
Inclusion of the latter in the above distributions causes only a small ($\le$
1 kpc) shift to lower values (for all the samples), but otherwise does not alter significantly the 
distributions.  Median and mean values (both with or excluding the $V$-band
diameters) are listed in Table~\ref{tab3}.
Note that we use {\em physical} (kpc) rather than apparent (arcsec) diameters. 
Although the two Seyfert samples have similar redshift distributions (see Figure 
~\ref{f7}), it is clear that the distribution of their diameters is not the
same: Seyfert 1s are overall {\it smaller} than Seyfert 2s. This is probably
related to the tendency we found earlier for Seyfert 1 galaxies to have fainter
disks compared to the Seyfert 2 galaxies in our sample.

\placefigure{f5}

\begin{figure}
\epsscale{1.0}
\plotone{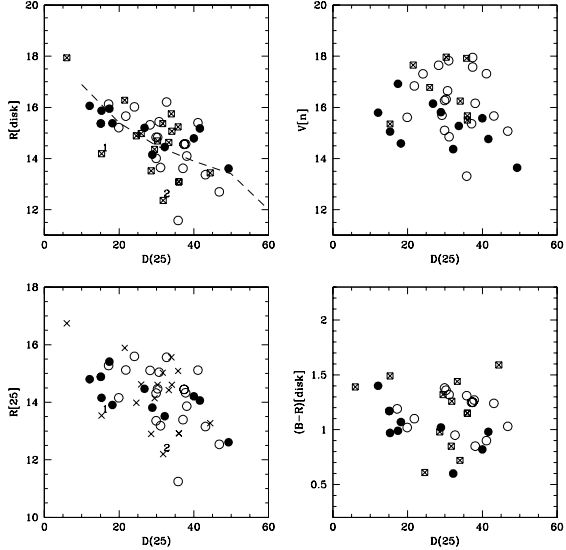}
\caption{Major axis diameters, corresponding to $\mu_{B}$=25 mag arcsec$^{-2}$ plotted against nuclear, disk and total magnitudes and disk colours. The dashed line indicates the expected correlation if disk luminosity scales with size (see text). The labels {\em 1} and {\em 2} indicate the types of Seyfert galaxies belonging to the Cold sample. \label{f6}}
\end{figure}

To investigate whether Seyfert 2 disks are brighter because they are larger or
because their surface brightness is higher, we tried to correlate major-axis 
diameters vs magnitudes for all scales; some of these plots are shown in Figure~\ref{f6}. We 
do find a correlation between diameters and disk magnitudes, for both the Seyfert
1 and Seyfert 2 samples. Assuming an exponential disk, the flux should be 
proportional to $h^{2}$ ($h$ being the disk scale length) or alternatively to 
$D^{2}$. Thus, for constant central surface brightness $\mu_{0}$, the disk 
luminosity will scale with disk size as \( M_1-M_2=-5 \log{D_1/D_2}\). In Figure~\ref{f6} we have plotted this relation (arbitrarily shifted on the y-axis) as 
a dashed line. The correlation appears to be tighter for Seyfert 1s, roughly 
following the expected correlation between disk size and luminosity. Seyfert 2s 
show larger scatter and a clear deviation from the above curve: at the 
large-$D_{25}$ end, for similar-sized hosts Seyfert 2s have brighter disks than 
Seyfert 1s; at the faint-disk end, for similar brightnesses Seyfert 2 disks tend 
to be larger. Thos indicates that Seyfert 2 disks tend to be larger 
{\em and} brighter than their Seyfert 1 counterparts. Total magnitudes show some 
correlation with host sizes, for both samples, but with larger scatter. 

Do {\em nuclear} magnitudes correlate with host dimensions? In the same figure we
show the $V_{nuclear}$ vs $D_{25}$ diagram. We do see a tendency for the brighter 
Seyfert 1 to reside in larger (and brighter) hosts, conclusion reached earlier
from our mag-mag plots. We don't find this tendency for Seyfert 2s; instead, as
we concluded earlier from our histograms, the Seyfert 2 nuclei tend to be fainter
than their Seyfert 1 counterparts for similar host sizes. To investigate the 
possible origin for the different host properties between the two samples we have
also plotted disk colours against sizes and luminosities.
There is a tendency for Seyfert 1s disks, when bluer to also be larger (and 
brighter), which could be simply consistent with a sequence of earlier to later 
type hosts. Such a correlation is clearly not seen for the Seyfert 2s, one 
possible interpretation for this being the larger amount of dust residing in 
Seyfert 2 disks, as we concluded earlier from their redder colours. 

\placefigure{f6}
 
In conclusion, in this section we have shown that there exist important differences between the
Warm Seyfert 1 and 2 galaxies. If differences in their nuclear optical
properties can be explained with larger torus obscuration of the Seyfert 2 
nuclei, this is not the case for their host properties: We find that the Seyfert
2 host disks tend to be larger, brighter and redder, result that is difficult
to accommodate within the simple orientation unification scheme for Seyferts.
Inspection of the contour maps presented in Paper I, shows a larger fraction of 
interacting systems among the Seyfert 2 galaxies as compared to the Seyfert 1s. 
In Paper V we will investigate the correlation between host sizes/luminosities 
and the interaction statistics and find that there is indeed a good correlation 
between these two for the Seyfert 2 galaxies.

\placefigure{f7}

\subsection{ Warm vs Cold Samples }

We have photometric information  in the $B,V,R$ bands for 14 out of 16 objects in
the Cold sample (see Table~\ref{tab2}). In this section we shall
investigate how the distributions of various photometric properties compare 
between the Warm Seyfert and the Cold samples.

In Figures~\ref{f1} and ~\ref{f2} we show the distributions 
of magnitudes and colours for the Warm (Seyfert 1 and 2) and Cold samples and in 
Table~\ref{tab3} we list the x[median and mean values for the Warm sample as a 
whole (and the two Seyfert subsamples) and for the Cold sample. 
As we will see in what follows, the Cold galaxies differ significantly from 
the Warm Seyfert 1s, while there is an overlap of properties between the
former and the Warm Seyfert 2s. Our main results are:

\subsection*{\it Magnitudes}

(i) Warm Seyferts tend to have brighter {\em nuclei} (Seyfert 1s by 1.5-2 mag,
Seyfert 2s by $\sim$0.5 mag) than their Cold counterparts. The Student's T-test 
shows significantly different means (significance levels 0.006-0.03) for the 
Seyfert 1 and Cold samples and the K-S test shows that the null hypothesis for 
matching distributions can be rejected at a significance level better than 95\%. 
On the other hand, there is a better overlap between the Warm Seyfert 2 and
Cold nuclear magnitudes (statistically significant for the $B$-band) at shorter
wavelengths, with a tendency for the Seyfert 2 nuclei to be brighter than their
Cold counterparts (statistically significant for the $R$-band magnitudes).

(ii) There is an overlap in disk magnitudes between the Seyfert 2 and Cold 
samples (the K-S test shows matching distributions at significance levels better 
than 99\% for the $B$ and $R$ magnitudes). The comparison is more ambiguous between 
Seyfert 1s and Cold galaxies, the latter tend to have brighter disks.  

(iii) There is a general overlap in the integrated (total) colours for all three 
(sub)samples, statistically significant between the Seyfert 2 and Cold samples, 
in the case of $B$ and $R$ magnitudes.

\begin{figure}
\epsscale{1.0}
\plotone{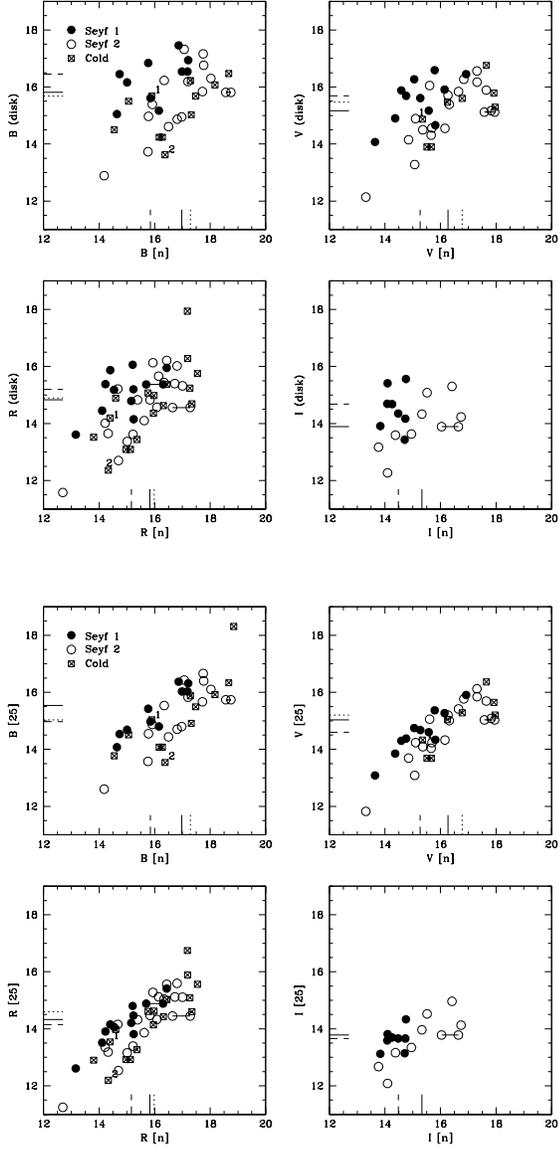}
\caption{Nuclear vs Disk and Nuclear vs Total mag-mag plots for the Warm Seyfert 1 and 2 type galaxies and for the Cold sample. Horizontal lines connect measures for the two nuclei of double nucleus systems. Labels 1 and 2 indicate Seyfert type 1 and 2 nuclei within the Cold sample galaxies. Bars indicate median values (to be compared with the mean values indicated in the histogram plots) for the three samples: dashed for Seyfert 1s, solid for Seyfert 2s and dotted for the Cold galaxies. \label{f3}}
\end{figure}

(iv) In the magnitude-magnitude plots of Figure~\ref{f3}, the nuclear 
magnitudes of Cold galaxies (indicated with crossed squares) scale with disk 
(and total) magnitudes as in the case of the Warm Seyfert 2 sample. There is a 
good overlap between the Seyfert 2 and Cold objects 
in all wavelengths, but there are some Seyfert 2s clustering together with the 
Seyfert 1s. This ``bimodality' of the Seyfert 2 sample is best seen in
the $R$ mag-mag plot (larger number of points). The faintest Cold object in the
$B$ and $R$ plots, is the southern member of the interacting system IRAS
09406+1018 (this is one of the systems for which both members are included in 
our analysis). Its isolation from the rest of the Cold sample might be indicating
that the IRAS source is instead the brighter, northern component.

(v) In Figure~\ref{f3} we indicate with {\bf 1} and 
{\bf 2} the Cold galaxies with observed Seyfert nuclear spectra. There are two 
more Seyfert 2s in our Cold sample but without photometric data. From the 
rest, five have \ion{H}{2}-region like spectra and seven have no spectroscopic 
information (see Table 2 of Paper I).
Although general conclusions cannot be drawn from a small number of objects,
it is interesting  that (i) the two Cold Seyferts occupy similar places in these 
diagrams as their Warm counterparts and that (ii) the \ion{H}{2} galaxies tend to 
overlap with the Seyfert 2 loci.

\subsection*{\it Colours}

(vi) The colour distributions for the three samples are shown
in Figure~\ref{f2} (there are only two Cold galaxies with simultaneous 
$B$ and $V$ band and none with $I$ band measures). In Table~\ref{tab3}, we list
indicatively ``median'' and ``mean'' $(B-V)$ colours.
The Cold sample seems to span the whole $(B-R)$ range 
covered by both the Warm Seyfert 1s and 2s.
The colour distributions are different between the Cold and Seyfert 1 samples
at all spatial scales (the former being redder) but there is an overlap in the
$(V-R)$ distributions between the Warm Seyfert 2 and Cold sample, that is 
statistically significant for the nuclear and total colours (K-S confidence
level better than 95\%). As we see in the colour-colour plots of Figure~\ref{f4} the Cold objects span indeed the whole extend of the Warm sample
colours. There is a good correlation between nuclear vs disk and nuclear vs total
colours, as in the case of Seyfert 2s.
Moreover, from the nuclear vs disk colour-colour diagram, we see that most Cold 
galaxies have negative colour gradients outwards, which we found to also be the 
case for the Warm Seyfert 2s.  

\subsection*{\it Host Sizes}

(vii) In Figure~\ref{f5} we show the distributions of major-axis 
diameters corresponding to $\mu_{25}$, for the Warm and the Cold samples. The 
distribution of Cold sample sizes is similar to that of Warm Seyfert 2s. On the 
mag vs $D_{25}$ plots (Figure~\ref{f6}), we see a tendency for larger Cold 
disks to be also brighter (same tendency for their total magnitudes), with a 
large scatter around the expected correlation. In this figure we also find a good
overlap between the Seyfert 2 and Cold samples and that the two Cold galaxies 
hosting a Seyfert nucleus have brighter nuclei {\em and} disks for similar host 
sizes, compared to the rest of the Cold sample.

\subsection{Main Conclusions}

The optical properties of our sample objects suggest {\em intrinsic} differences 
between the IR-Warm Seyfert 1 and 2 hosts, as well as, a larger degree of nuclear
obscuration in Seyfert 2s. Moreover, the Cold and Warm Seyfert 2 samples show a 
significant overlap of properties. In particular, we find that:

(i) Seyfert 1 nuclei are brighter than their Seyfert 2 counterparts, this effect
increasing at shorter wavelengths. Seyfert 1 nuclei are also bluer than Seyfert 
2s for similar disk colours, this being probably the reason for the opposite 
colour gradients that we found in Seyfert 1s and Seyfert 2s (red/blue outwards, 
respectively). We can interpret these results as indicating either intrinsically 
brighter Seyfert 1 nuclei or larger obscuration of the Seyfert 2 nuclei. 

(ii) Seyfert 2 disks are larger and brighter compared to their Seyfert 1 
counterparts while the total magnitudes are similar for the two samples. The 
latter indicates that the Seyfert 1 optical properties are dominated by their 
nuclei and that Seyfert 1s are more compact than Seyfert 2s.

(iii) Extinction affects both the nuclei and hosts of Warm Seyfert 2s, that
tend to be overall more dusty that Seyfert 1s (redder nuclear/disk/total colours).

(iv) There are no pronounced differences in the optical properties of the Warm 
(taken as a whole) and the Cold samples, except for a tendency of the Cold 
galaxies to be fainter and have redder nuclei. However, when comparing the Cold 
sample with the Seyfert 1 and 2 samples separately, we find that Cold galaxies 
are significantly different from the Warm Seyfert 1 galaxies as a sample, while 
there is a significant overlap in the optical properties (magnitudes, sizes and 
colours) of the Cold and Seyfert 2 samples.

\section{Optical vs IR properties}

\subsection{IR Fluxes and Colours}

\subsection*{\it Redshifts and IR Luminosities}
 X[
In Figure~\ref{f7} we show the distributions of 25 and 60 $\mu$m 
luminosities and redshifts for the Warm Seyfert 1 and 2 and the Cold galaxies, 
with photometric information. 

(i) As we showed in Paper I for the totality of the objects, all three samples 
occupy a similar range in redshifts. The only difference is the inclusion in
the Warm sample of a few high-luminosity high-z Seyfert 2 galaxies (see Paper I)
in order to investigate the observed paucity of high-luminosity narrow-line AGNs.

(ii) A comparison of the distributions of the 25 $\mu$m luminosities $L_{25}$ 
shows significant differences between the three (sub)samples.
The Warm Seyfert 1s span a very narrow range in $L_{25}$, compared to the other
two samples. Seyfert 2s tend to have larger 25 $\mu$m luminosities than Seyfert 
1s and, as expected, the Warm sample galaxies are brighter 25 $\mu$m emitters 
than 
the Cold galaxies. The Student's T-test shows statistically significant different
means between the Warm and Cold samples (significance level $\leq$0.009) and the 
K-S test shows that the null hypothesis that any of the Warm Seyfert and the Cold
samples are drawn from the same population of 25 $\mu$m luminosities can be 
rejected at a confidence level better than 96\%. 

(iii) The three samples span a similar range in 60 $\mu$m luminosities. The Warm 
Seyfert 2 and the Cold galaxies have comparable distributions while the Warm
Seyfert 1 distribution is shifted to somewhat lower luminosities
(but with less than 95\% statistical significance according to the K-S test).

How can these results be understood?
\cite{fadda98} have found larger 25-100 $\mu$m luminosities for type 2
nuclei compared to their type 1 counterparts, using  an extended sample of 
Seyfert galaxies. They also found a similarly strong effect among UV-excess 
Seyferts but much weaker among IR-selected Seyferts. They concluded that it might
be a selection effect, due to preferentially including in the optically selected 
samples Seyfert 2s with strong (mainly disk) star formation. Our samples include 
only Seyferts that were IR-selected but we find the same effect. 
In Paper I we have discussed a possible selection 
effect towards more powerful type 2 AGNs in our Warm sample, introduced by the optical
spectroscopic classification of \cite{grijp92}. If this is true, it could 
explain the larger IR luminosities of these objects. However, in the same paper
we showed that, on the basis of the IR colour selection criteria, it is
unlikely that any low 
luminosity AGNs are included in the mid-IR Warm Seyfert samples, this being 
true for both Seyfert types. In turn, we suggest that the larger IR luminosities 
found for Seyfert 2s both in our sample and in those of Fadda \etal might indicate an 
{\em intrinsic} difference between the two classes of Seyferts.
Possibilities are: 
(i) the mixed nature of Seyfert 2s (AGN plus starbursts), in which case the 
excess 25 $\mu$m and 60 $\mu$m emission is associated with excess star formation 
in Seyfert 2s, (ii) the larger (hot) dust content in Seyfert 2s, maybe related to
recent interactions/mergers in these objects or (iii) differing contributions of 
the {\em nuclear} components of the 25 $\mu$m emission (the AGN warming the 
dust more efficiently in Seyfert 2s, for some reason). Any one of these alternatives 
are difficult to reconcile with the simple torus obscuration model for Seyferts.

\begin{figure}
\epsscale{1.0}
\plotone{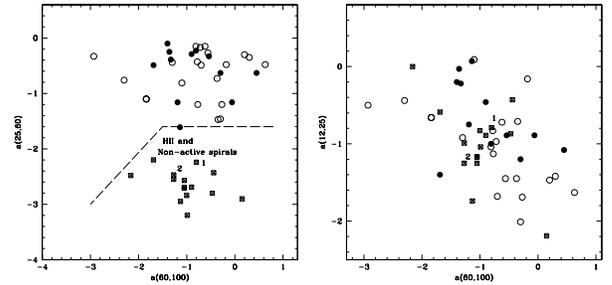}
\caption{Mid- and far-IR colour-colour diagrams. Filled/open circles indicate Warm Seyfert 1s and 2s, respectively and crossed squares indicate the Cold galaxies. The labels 1 and 2 indicate the respective types of Seyfert nuclei within the Cold sample. The dashed line on the left panel approximately divides the AGN and normal galaxy regions according to the mid-IR colour selection criterion of \cite{grijp87}. \label{f8}}
\end{figure}

\placefigure{f8}

The modeling of thick dusty tori in Seyferts has shown that Seyfert 1s are 
expected to be stronger near- and mid-IR emitters (face-on tori) than Seyfert 2s,
but to have similar far-IR luminosities. However, if the 25 $\mu$m emission is 
mainly of nuclear origin, our result suggests that the tori are optically thin 
already at this wavelength.
In Figure~\ref{f8} we plot the $\alpha_{(25,60)}$ vs $\alpha_{(60,100)}$
colour-colour diagram for the Warm and Cold galaxies discussed in this paper
(with photometric information). This plot illustrates the 
basic selection criterion $\alpha_{(25,60)}$ between our Warm and Cold 
samples.
The $\alpha_{(60,100)}$ colour is commonly used as an indicator of dust heated
by current star formation but it can also be affected by the cooler 
interstellar radiation field. For this reason, the $\alpha_{(12,25)}$ is
considered to be a better indicator of warm dust associated with star
formation or/and an active nucleus. Thus, in Figure~\ref{f8} we also show 
the $\alpha_{(12,25)}$ vs $\alpha_{(60,100)}$ colour-colour plot. Here, Seyfert 2s tend to 
have
cooler (redder) 12-25 $\mu$m colours than Seyfert 1s. This could be either the 
effect of the dusty torus being optically thick at $\lambda<$25 $\mu$m (which we
excluded earlier) or that Seyfert 2s are stronger 25 $\mu$m IR emitters compared
to Seyfert 1s, in favour of which we argued earlier.
On the same colour-colour plot we find a tendency for cooler 12-25 $\mu$m 
towards warmer 60-100 $\mu$m colours, in particular for Seyfert 2s that tend to
occupy the lower right part of the plot. Model predictions (\eg \cite{xu89};
\cite{mazzarella91}) for non-Seyfert galaxies indicate that the relative 
importance and temperature of the warm dust component, related to star formation,
increases as we progress from the upper left to the lower right corners of this 
diagram. Thus, the relative positions of our Warm subsamples suggest that Seyfert 2 
galaxies have a larger fraction of warm dust related to increased star formation,
compared to Seyfert 1s, while the generally warmer 12/25 colours of the latter
indicate larger contribution of the nuclear component in these objects.  

The distribution of IR luminosities for the Cold sample, shown in Figure~\ref{f7}, is shifted towards smaller 25 $\mu$m and larger 60 $\mu$m emission compared to the Warm sample and 
is easily understood in the context of our mid-IR colour criterion.
In the $\alpha_{(25,60)}$ vs $\alpha_{(60,100)}$ colour-colour plot of Figure~\ref{f8}
the Warm and Cold samples are well separated by their $\alpha_{(25,60)}$ 
colours (as was postulated by our main selection criterion), while they
span similar ranges in $\alpha_{(60,100)}$ colours. On the $\alpha_{(12,25)}$ vs $\alpha_{(60,100)}$ 
colour-colour plot, Cold galaxies tend to segregate around intermediate values.
Their colder 60-100 $\mu$m colours compared to Warm Seyferts, for similar 
12-25 $\mu$m colours, indicate a significant contribution of colder dust at
100 $\mu$m in these objects. 

The narrow range of Seyfert 1 25 $\mu$m luminosities that we found earlier, is 
more difficult to understand. One possible explanation for this would be their 
differing dust properties compared to the Seyfert 2s in our sample; if, for 
instance, we assume that in Seyfert 1s the AGN is the main source of mid-IR 
emission through thermal dust re-radiation, at some level of nuclear power and 
above, the small dust grains (responsible for the emission at this spectral 
range) tend to be destroyed. In Seyfert 2s this effect is less noticeable either 
because the central AGN is intrinsically fainter or because of the mixed nature 
(AGN+starburst) of their 25 $\mu$m emission.

In conclusion, we can reconcile the differences of IR fluxes and colours 
between Seyfert 1 and 2 types, considering the alternatives (i) and (ii)
mentioned earlier, \ie mixed nature (AGN+starbursts) and larger dust content
for the Seyfert 2 galaxies. This however would lead to problems the simple obscuration picture, where 
the only difference between the Seyfert types should be their (projected) 
orientation on the sky.

\subsection{IR Loudness Indices}

\placetable{tab4}

\begin{figure}
\epsscale{0.85}
\plotone{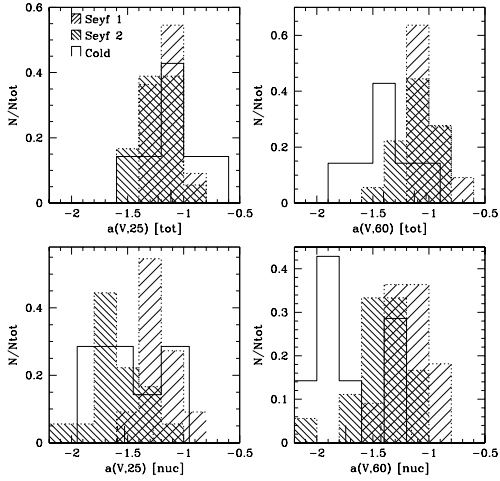}
\caption{Fractional distributions of the  $\alpha_{(V,25)}$ and $\alpha_{(V,60)}$ IR-loudness indices for the Warm Seyfert 1 and 2 and the Cold galaxies, with photometric information. The vertical bars on the lower x-axes indicate mean values for the two subsamples (dotted for Seyfert 1s, dashed for Seyfert 2s and solid for Cold objects). \label{f9}}
\end{figure}

The IR-loudness is defined by the optical-to-IR indices 
\( \alpha_{(V,25)}=0.60 \log{\frac{f_V}{f_{25}}} \)
and \( \alpha_{V,60}=0.49 \log{\frac{f_V}{f_{60}}} \)
where {\em f} denotes flux densities in units of Jy.
The IR-loudness is primarily a measure of extinction and has been used as 
a warm dust indicator together with the $\alpha_{(60,100)}$ colour index.
Since we do not know the spatial extent of the mid-IR emission, we
calculated the IR-loudness indices for both the nuclear and total $V$ 
magnitudes and display their distributions in Figure~\ref{f9} and their 
mean values in Table~\ref{tab4}.
Our main conclusions are:

\placefigure{f9}

(i) Seyfert 2 galaxies tend to be IR-louder than Seyfert 1s in all scales 
the difference being more pronounced and statistically significant for the
nuclear indices and the 25 $\mu$m IR emission: the Student's T-test shows that
their means differ at significance level $\sim$0.002 and the 
K-S test shows that the null hypothesis that the two samples have matching 
distributions of nuclear IR-loudness can be rejected at a significance level 
better than 95\%. This result can be understood in terms of the larger optical
extinction (in particular nuclear) and the stronger IR emission (in particular
at 25 $\mu$m) found for Seyfert 2s.

(ii) The Cold galaxies are clearly IR-louder as far as the $\alpha_{(V,60)}$ index is 
concerned, at all spatial scales. This is primarily due to the excess 60 $\mu$m 
emission that characterizes the Cold sample (discussed earlier).
The shift is more pronounced for the nuclear quantities, due to their fainter
nuclear magnitudes (see Section 2.2). The Student's T-test shows that the mean 
values of $\alpha_{(V,60)}$ differ significantly between the Warm Seyfert 1 and 2 and 
the Cold samples (significance $\leq$0.005) and the K-S test shows that the null 
hypothesis of matching distributions between either of the Warm and the Cold 
sample can be rejected at a significance level better than 95\%. 

(iii) There is a good overlap in the distributions of $\alpha_{(V,25)}$ between the Warm
and Cold samples. This is probably due to the competing effects of fainter $V$ 
magnitudes for the Cold galaxies and larger 25 $\mu$m excess for the Warm sample.

\subsection{IR vs Optical Luminosities, Ionization and Host Sizes}

\subsection*{\it $L_{IR}$ vs $L_{Optical}$}

\begin{figure}
\epsscale{0.85}
\plotone{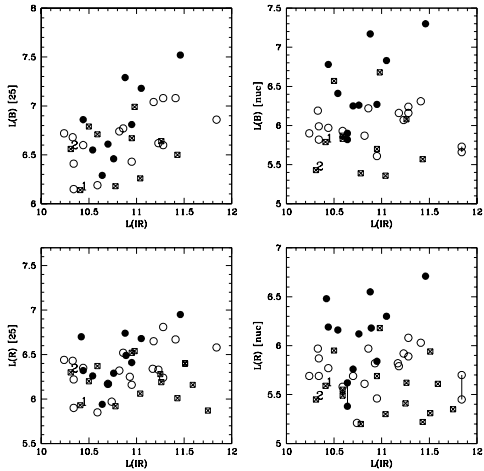}
\caption{Far-IR (60-100 $\mu$m) vs optical (nuclear and total) luminosities for the Warm Seyfert 1 and 2 subsamples (filled/empty circles, respectively) and for the Cold sample galaxies (crossed squares). Solid lines connect points corresponding to the two nuclei of double nucleus systems. The labels {\em 1} and {\em 2} indicate the respective Seyfert types of Cold sample galaxies. Luminosities are expressed in solar units \Lsun. \label{f10}}
\end{figure}

\placefigure{f10}

\begin{figure}
\epsscale{1.0}
\plotone{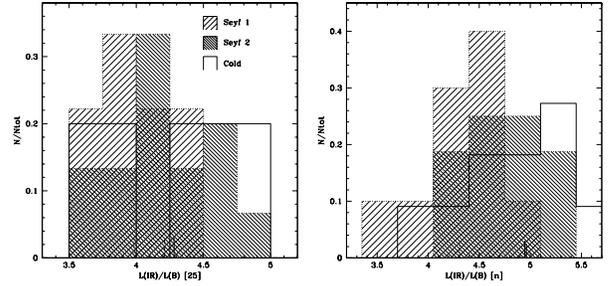}
\caption{Fractional distributions of the \( \frac{L_{IR}}{L_{B}} \) ratio between the Warm Seyfert 1 and 2 and the Cold samples. The vertical bars on the lower x-axes indicate mean values for each subsample (dotted for Seyfert 1s, dashed for Seyfert 2s and solid for Cold objects). \label{f11}}
\end{figure}

\placefigure{f11}

For a normal (non IR-luminous) galaxy, the far-IR luminosity (60-100 $\mu$m)
$L_{IR}$ scales with blue luminosity $L_{B}$ and galaxy size, because the IR 
traces dust heated by the older stellar population. The ratio 
\( \frac{L_{IR}}{L_{B}} \) tends to increase (there is an ``excess'') when active
star formation or an active nucleus are present.
It follows that this ratio has often been used as a measure of the young/older
stellar population dominance. However, it is not a very accurate indicator, if
the IR and blue luminosities arise from different size areas; moreover, dust
extinction would artificially increase this ratio. Having those in mind,
in Figure~\ref{f10} we plot $L_{IR}$ vs $L_{B}$  and $L_{R}$, nuclear 
and total, for the Warm and Cold samples. The main conclusions are:

(i) There is no obvious correlation between far-IR and 
nuclear optical luminosities for either of the Warm Seyfert samples.
Seyfert 2s span a large range in $L_{IR}$ for a relatively narrow range of 
{\em nuclear} optical magnitudes. The opposite trend is seen for Seyfert 1s (with
the exception of one object).  

(ii) We find a trend for $L_{IR}$ to correlate with {\em total} optical 
luminosities for both Seyfert subsamples, statistically significant for the
Seyfert 2s (correlation coefficient $\sim$0.5 with significance of the Spearman's
rank test $\sim$0.03).
The lack of any correlation between $L_{IR}$ and nuclear optical luminosities
seen above, indicates that the far-IR emission 
in Seyferts is dominated by warm dust in their host galaxy disk. 

(iii) There is an overlap between Cold and Warm Seyfert 2 galaxies in these 
diagrams, but no obvious correlation between far-IR and optical luminosities for
the Cold galaxies. This probably indicates important dust extinction at all 
spatial scales in the latter, that affects their optical luminosities.
The two Cold galaxies with Seyfert nuclei are among the less IR bright among both
the Cold and Warm samples.

(iv) At large IR luminosities, Seyfert 2s and Cold galaxies tend to be optically
fainter compared to Seyfert 1s for similar $L_{IR}$. If this is the result of 
larger optical extinction, it would mean that the far-IR emission in the former 
objects scales with their dust content. This in turn, means that the
 \( \frac{L_{IR}}{L_{B}} \) ratio is also affected by dust extinction, especially
at high IR luminosities ($L_{IR}\ge$10$^{11}\Lsun$).

(v) Having this in mind, we now compare the distributions of the 
\( \frac{L_{IR}}{L_{B}} \) ratio between the Warm Seyfert 1 and 2 and the Cold 
galaxies in Figure~\ref{f11}, calculated for both nuclear and total 
$B$-band magnitudes. 
We find indeed a larger \( \frac{L_{IR}}{L_B} \) ratio (excess IR to optical 
emission) for the Seyfert 2s compared to the Seyfert 1s, in particular for 
their {\em nuclear} blue luminosities (the K-S test shows that the two 
distributions are different at the 95\% significance level). The Cold galaxies
have similar \( \frac{L_{IR}}{L_B} \) distributions to those of the Warm Seyfert
2 sample (matching at the 97\% significance level). 

\subsection*{\it $L_{25}$ vs $L_{Optical}$}

\begin{figure}
\epsscale{1.0}
\plotone{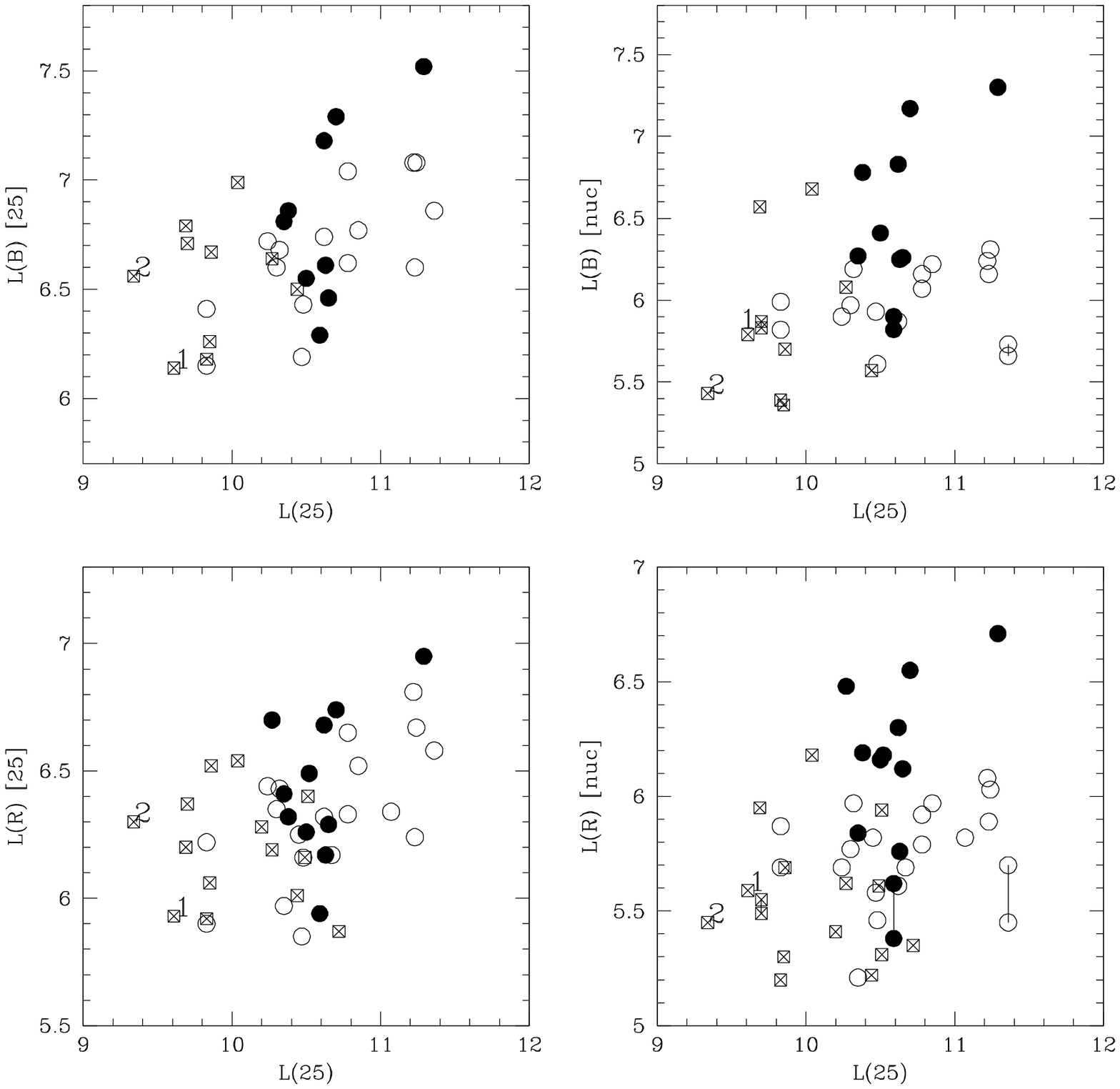}
\caption{Mid-IR (25 $\mu$m) vs optical (nuclear and total) luminosities for the Warm Seyfert 1 and 2  (filled/empty circles, respectively) and the Cold samples (crossed squares). Solid lines connect points corresponding to the two nuclei of double nucleus systems. The labels {\em 1} and {\em 2} indicate the respective Seyfert types of Cold sample galaxies. Luminosities are expressed in solar units \Lsun. \label{f12}}
\end{figure}

In Figure~\ref{f12} we plot the $L_{25}$ vs $L_{B}$ and $L_{R}$, 
nuclear and total. 

(i) As we saw earlier (Figure~\ref{f7}), Seyfert 1 nuclei occupy a very 
narrow strip in all diagrams, that is, they have similar 25 $\mu$m  
luminosities for a wide range of optical luminosities (there is only one object,
IRAS 00509+1225 (I18), that is detached from the rest of the sample towards 
higher $L_{25}$). Consequently, Seyfert 1s show no correlation between 25 $\mu$m 
and optical luminosities, at all wavelengths and spatial scales.

(ii) Seyfert 2 galaxies span a large range in $L_{25}$. Their 25 $\mu$m
luminosities show a definite correlation with {\em total} (and disk) optical 
luminosities (correlation coefficient 0.7 with significance 0.004). Such a 
correlation is however not clear for their nuclear optical luminosities. This 
indicates that at least some of the 25 $\mu$m emission in Seyfert 2s originates 
in their {\em host disks} while their nuclear regions remain highly obscured. The
latter is further supported by the fact that in Figure~\ref{f12} the 
Seyfert 2 nuclei are $\sim$1 order of magnitude fainter in the optical than 
their Seyfert 1 counterparts, for the whole range of 25 $\mu$m luminosities.
(The two galaxies detached somewhat from the Seyfert 2 sample by their lower 
$L_{25}$, are the two members of the interacting triple system IRAS 11298+5313 
(I283)).

(iii) The Cold galaxies, occupy the lower left region in all the above diagrams, 
that is, towards lower 25 $\mu$m and optical luminosities. They show large 
scatter and no obvious correlations. The two Cold Seyferts show the lowest 
$L_{25}$ luminosities of all samples.

\placefigure{f12}

\subsection*{\it IR luminosities vs Ionization}

\begin{figure}

\epsscale{1.0}
\plotone{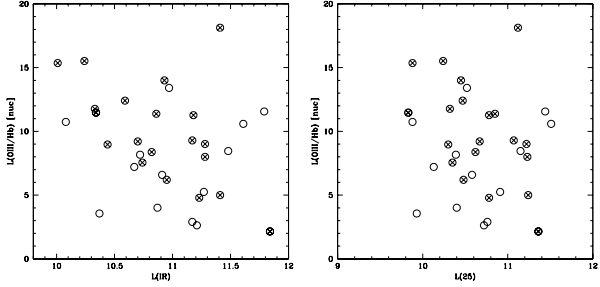}
\caption{Mid- and far-IR luminosities against the ionization ratio [\ion{O}{3}]$_{5007}/H_{\beta}$, for the totality of our Warm Seyfert 2 subsample. The crossed circles indicate objects for which we also have photometric information. Luminosities are in solar units \Lsun. \label{f13}}
\end{figure}

We have considered possible correlations between IR luminosities ($L_{IR}$,
$L_{25}$) and optical emission line fluxes and ratios ($H_{\alpha}$,
[\ion{O}{3}]$_{5007}$, [\ion{O}{3}]$_{5007}/H_{\beta}$), using 
spectroscopic 
(predominantly nuclear) data obtained by \cite{grijp92}. We can use those
data only for the Warm Seyfert 2 sample, because the line fluxes for Seyfert 1s include
contributions from both the broad and narrow Balmer components.
As we have shown earlier, the mid- and far-IR emission in Seyfert 2s is probably 
dominated by dust in their host disks, consequently we do not expect any strong 
correlation between {\em nuclear} parameters and IR quantities. This is indeed 
the case as far as emission line fluxes are concerned, a result that agrees with 
the conclusions reached earlier by \cite{keel94} for the totality of the Warm 
sample. However we find a tendency for {\em anti-correlation} between both the
far- and mid-IR fluxes and the ionization ratio 
[\ion{O}{3}]$_{5007}/H_{\beta}$, which is shown in Figure~\ref{f13}. 
If true, this is an intriguing result and contradicts the correlation that 
\cite{keel94} found between the same ionization ratio and the IR-warmth 
$\alpha_{(25,60)}$ index (for which we find no evidence in our subsample) and their 
conclusion,  that both quantities are dominated by the AGN. 

\placefigure{f13}

\subsection*{\it IR Luminosities vs Host Sizes}
 
\begin{figure}
\epsscale{1.0}
\plotone{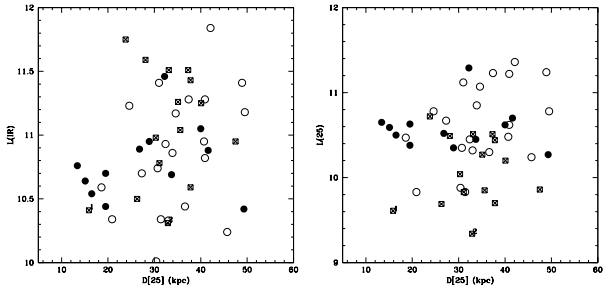}
\caption{Mid- and far-IR luminosities against 25 mag arcsec$^{-2}$ diameters, for the Warm and Cold samples. Filled/open circles represent Warm Seyfert 1/2 galaxies and crossed squares Cold galaxies. The labels 1 and 2 indicate the corresponding nuclear types of Cold Seyferts. Luminosities are in solar units \Lsun. \label{f14}}
\end{figure}

In Figure~\ref{f14} we check the dependence of IR luminosity 
on galaxy size, represented by the major axis length at $\mu_{B}$=25 
mag arcsec$^{-2}$. 

(i) There seems to exist a correlation between size and far-IR
emission, for both Seyfert 1 and 2 types, but with large scatter (thus, not
statistically significant). The two Seyfert 1 galaxies that deviate the most from
this relation with over/underestimated 
diameters for their IR luminosities are IRAS 05136-0012 (I171) and IRAS
00509+1225 (I18), respectively. The $B$-band images of both objects are not 
very 
deep, consequently the extrapolation of their surface brightness profiles and
their estimated diameters are less accurate. The same holds for the two Seyfert 
2 galaxies with apparently underestimated sizes, for their IR luminosities 
(IRAS 09305-8408 (I254) and IRAS 15599+0206 (I392)). The Seyfert 2 galaxy that 
deviates from the correlation, with very large diameter for its IR luminosity is 
IRAS 02580-1136 (I67), due to its bright southern tidal arm.

(ii) The Cold sample shows a similar scaling between far-IR luminosity and host
galaxy size, but this breaks down at the high end of $L_{IR}$, for four galaxies
(IRAS 04454-4838, 23179-6929, 04530-3850 and 09406+1018) that are all members of
closely interacting systems, tentatively suggesting that these objects tend to
have larger far-IR excesses.

(iii) When plotting the 25 $\mu$m emission vs galaxy size (same figure), we find 
no correlation for the Seyfert 1 galaxies, as expected if their 25 $\mu$m emission
is coming from the nuclear region, but a correlation (0.05 significance level)
is seen for the Seyfert 2s.

\placefigure{f14}

\subsection{IR Properties vs Optical Colours}

\subsection*{\it IR Luminosities}

\begin{figure}
\epsscale{1.0}
\plotone{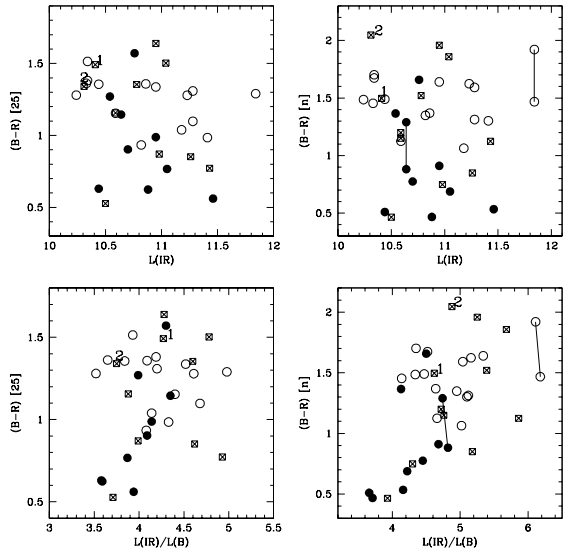}
\caption{Far-IR (25-60 $\mu$m) luminosities and excess ratio \( \frac{L_{IR}}{L_{B}} \) vs optical (nuclear and total) colours for the Warm Seyfert 1 and 2 subsamples (filled/empty circles, respectively) and the Cold sample galaxies (crossed squares). Solid lines connect points corresponding to the two nuclei of double nucleus systems. The labels {\em 1} and {\em 2} indicate the respective Seyfert types of Cold sample galaxies. Luminosities are in solar units \Lsun. \label{f15}}
\end{figure}

In Figure~\ref{f15} we are plotting the $L_{IR}$ luminosity and excess 
ratio \( \frac{L_{IR}}{L_{B}} \) vs optical colours, nuclear and total.

(i) We find large scatter in the plots involving $L_{IR}$ and a tendency for 
Seyfert 1 \( \frac{L_{IR}}{L_{B}} \) to correlate with redder colours, which can
be  explained if $L_{IR}$ scales with galaxian light and the optical colours 
are affected by dust obscuration. No correlations are found for the Seyfert 2 or 
the Cold samples. 

(ii) Seyfert type 1 and 2 galaxies occupy different regions in the above plots:
from bluer colours and smaller IR excess for Seyfert 1s, to redder colours
and larger IR excess for Seyfert 2s. As we discussed earlier this is 
probably due to larger dust extinction in the latter.

\placefigure{f15}

\subsection*{$\alpha_{(25,60)}$}

\begin{figure}
\epsscale{1.0}
\plotone{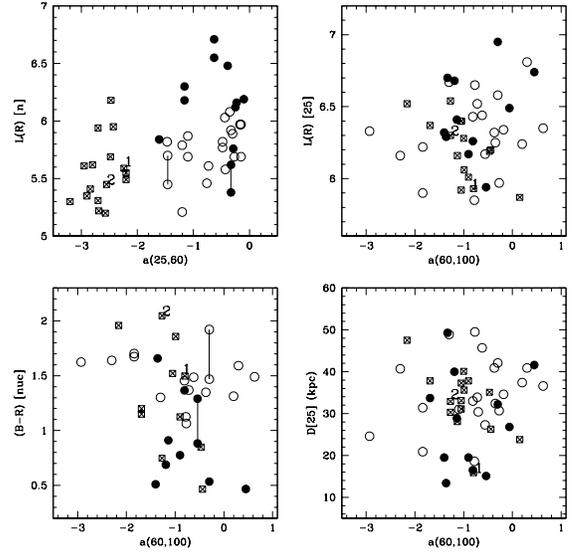}
\caption{IR colour indices vs optical luminosities, nuclear ($L_{R}(n)$) and total ($L_{R}(25)$), nuclear $(B-R)$ colours and host sizes $D_{25}$, for the Warm Seyfert 1 and 2 subsamples (filled/empty circles, respectively) and the Cold sample galaxies (crossed squares). Solid lines connect points corresponding to the two nuclei of double nucleus systems. The labels {\em 1} and {\em 2} indicate the respective Seyfert types of Cold sample galaxies. Luminosities are in solar units \Lsun. \label{f16}}
\end{figure}

\placefigure{f16}

There is no significant correlation between mid-IR warmness, expressed by this 
colour index and optical properties (luminosities or colours), except for a 
tendency for Seyfert 2 {\em nuclear} luminosities to have warmer (bluer)
25-60 $\mu$m colours, shown in Figure~\ref{f16}. Mid-IR warmness does
not correlate with either $L_{IR}/L_{B}$, galaxy size, or with any of the 
spectral (predominantly nuclear) properties of Seyfert 2s ($H_{\alpha}$ emission,
ionization ratio [\ion{O}{3}]$_{5007}/H_{\beta}$, dust reddening 
$H_{\alpha}/H_{\beta}$).

\subsection*{$\alpha_{(60,100)}$}

We find no correlation between $\alpha_{(60,100)}$ and optical properties for the
Warm sample. On the other hand, the Cold sample shows a tight {\em 
anti-correlation} between far-IR colours and $R$-band total luminosities or host 
sizes. We also find warmer far-IR colours to scale with bluer optical colours 
(Figure~\ref{f16}), which could be explained if both are due to star 
formation. There is no significant correlation between $\alpha_{(60,100)}$ and IR excess
or optical spectral properties for any of the samples.

\subsection*{$\alpha_{(12,100)}$}

\begin{figure}
\epsscale{1.0}
\plotfiddle{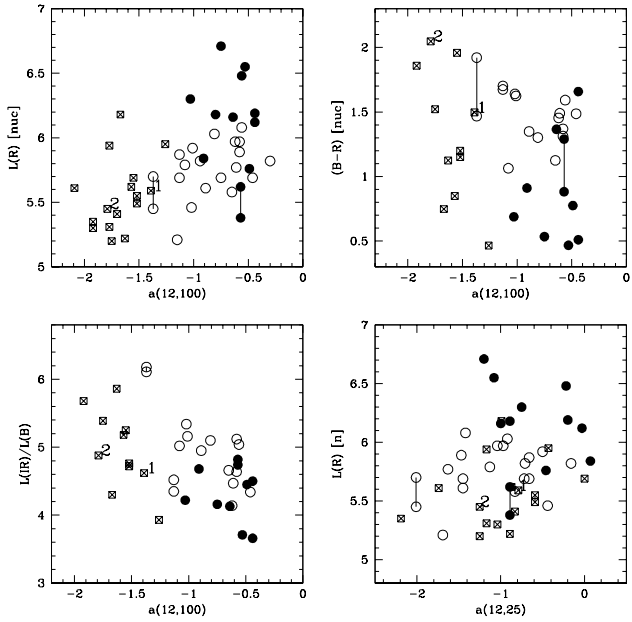}{400pt}{0}{45}{45}{-150}{-70}
\caption{IR colour indices vs nuclear luminosities ($L_{R}(n)$) and $(B-R)$ colours and IR/optical excess ($L_{IR}/L_{B}$), for the Warm Seyfert 1 and 2 subsamples (filled/empty circles, respectively) and the Cold sample galaxies (crossed squares). Solid lines connect points corresponding to the two nuclei of double nucleus systems. The labels {\em 1} and {\em 2} indicate the respective Seyfert types of Cold sample galaxies. Luminosities are in solar units \Lsun. \label{f17}}
\end{figure}

\placefigure{f17}

In Figure~\ref{f17} we plot the 12-100 $\mu$m colour indices against 
various optical properties. The Warm and Cold samples are well-separated by 
this colour index, indicating that
$\alpha_{(12,100)}$ is a fairly good discriminator between AGN-dominated and 
normal/starburst-dominated galaxies. We find a 
correlation between warmer 12-100 $\mu$m colours and {\em nuclear} optical 
luminosities for the Warm Seyfert 2 sample. There is also a tendency for cooler 
$\alpha_{(12,100)}$ to correlate with nuclear extinction as measured by the 
\( \frac{H_{\alpha}}{H_{\beta}} \) ratio in Seyfert 2s. No significant correlations are
found between $\alpha_{(12,100)}$ and any optical properties, for the Warm Seyfert 1 
galaxies. In fact, for these objects the $\alpha_{(12,100)}$ colour spans a relatively
narrow range for a large range in optical nuclear luminosities.
A tendency exists for $\alpha_{(12,100)}$ to correlate with nuclear optical luminosities
for the Cold sample, as in the case of Seyfert 2s. We also find a trend for 
cooler $\alpha_{(12,100)}$ with far-IR excess ratio $L_{IR}/L_{B}$. 

These results indicate the nuclear origin of the 12 $\mu$m emission in the Warm
Seyfert 2 galaxies. On the plots of Figure~\ref{f17} there is a continuous
transition from cooler to warmer 12-100 $\mu$m colours, fainter to brighter 
optical nuclear magnitudes, redder to bluer optical nuclear colours and larger to
smaller far-IR excess, as we progress from the Warm Seyfert 1, to the Warm 
Seyfert 2, to the Cold galaxies loci.

\subsection*{$\alpha_{(12,25)}$}
   
Plots involving this colour show large scatter. No significant correlations were 
found between the $\alpha_{(12,25)}$ colour index and any optical properties for any
of the samples (\eg Figure~\ref{f17}).

\subsection{IR Loudness vs Optical Properties}

\begin{figure}
\epsscale{1.15}
\plotfiddle{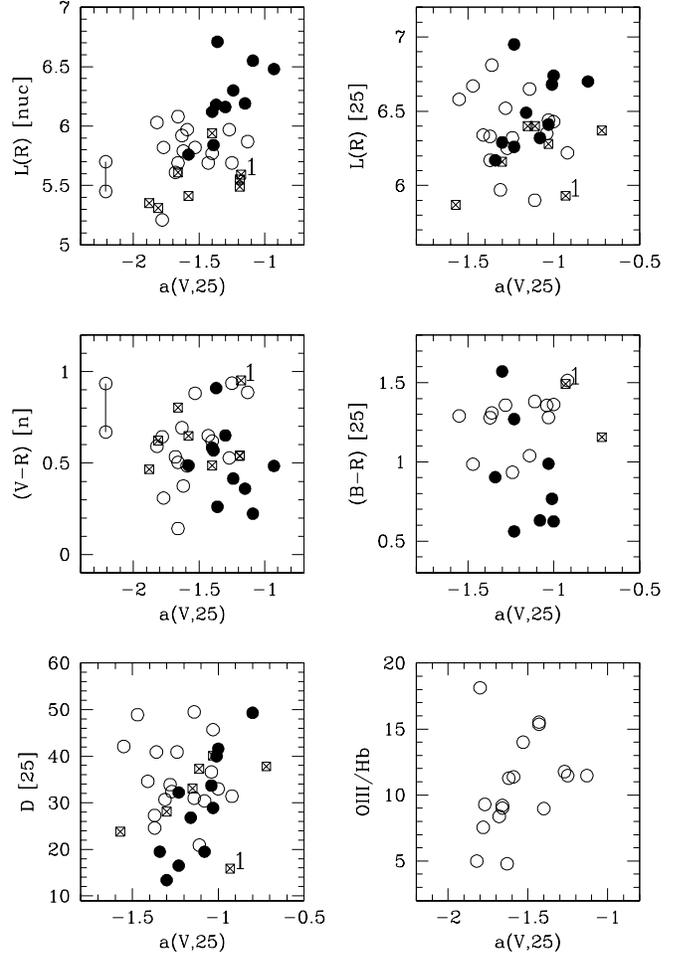}{400pt}{0}{70}{70}{-220}{-120}
\caption{IR-loudness indices vs optical luminosities $L_{R}$ and colours, nuclear and total, host size ($D_{25}$) and ionization ratio [\ion{O}{3}]/$H_{\beta}$, for the Warm Seyfert 1 and 2 subsamples (filled/empty circles, respectively) and the Cold sample galaxies (crossed squares). Solid lines connect points corresponding to the two nuclei of double nucleus systems. The labels {\em 1} and {\em 2} indicate the respective Seyfert types of Cold sample galaxies. Luminosities are expressed in solar units \Lsun. \label{f18}}
\end{figure}

We have searched for trends/correlations between the IR-loudness indices, 
$\alpha_{(V,25)}$ and $\alpha_{(V,60)}$, and various optical properties 
(Figure~\ref{f18}). We find: 

(i) IR-louder Seyfert 1s tend to be optically fainter, the correlation holds for 
both nuclear and total magnitudes. This correlation is expected if the 
IR-loudness is primarily a measure of dust extinction. This seems to be confirmed
by the tendency of IR-louder Seyfert 1s to have redder (nuclear and total) 
optical colours.

(ii) We find no such correlations for Seyfert 2s and even an opposite trend is
found for IR-loudness to scale with {\em total} optical luminosities
and bluer nuclear optical colours. We had seen earlier that the 25 (and 60)
$\mu$m emission is Seyfert 2s correlates with optical luminosities and this is 
the most obvious reason for the IR-loudness correlations that we find here. We 
thus conclude that the {\em total} IR-loudness in Seyfert 2s is indicative of 
their IR emission rather than dust extinction. We can check these ideas also by 
plotting the IR-loudness indices vs nuclear reddening (as expressed by the 
spectral $H_{\alpha}/H_{\beta}$ ratio) for Seyfert 2s. We find no significant 
correlation. 

(iii) We have fewer points for the Cold sample and no clear trend could be 
defined from the above diagrams. On the IR-loudness vs nuclear optical
luminosity diagram, there is a more or less continuous transition from the Cold 
and Seyfert 2 galaxies towards the Seyfert 1s as we progress from optically 
fainter and IR-louder towards brighter and less IR-loud nuclei.

All the trends/correlations described above hold for both IR-loudness indices but are 
better defined for the $\alpha_{(V,25)}$ index.

(iv) As expected, the $L_{IR}/L_{B}$ ratio is well correlated with the 
IR-loudness indices for all samples (not shown here), the correlation being 
particularly tight for the $\alpha_{(V,60)}$ index.

(v) In Figure~\ref{f18} we also plot the IR-loudness indices vs host 
galaxy size. We find a good correlation for Seyfert 1s in the sense, IR-louder
objects tend to also be more compact. This is probably a consequence of the good
correlation found earlier between optical {\em luminosities} and galaxy size. 
The same trend seems to exist for the Cold sample, although we have a smaller
number of points. In the case of Seyfert 2s there is no clear correlation, which 
is expected since as we showed earlier both optical and IR luminosities correlate
with galaxy size and between them, for these objects.

(vi) We find an anti-correlation between IR-loudness and the ionization ratio
[\ion{O}{3}]$_{5007}/H_{\beta}$ for Seyfert 2s, which is expected since a 
similar anti-correlation was found earlier between this ratio and IR luminosity.

\subsection{Main Conclusions}

Comparison between the optical and IR properties of our samples lead to the
following conclusions:

(i) Warm Seyfert 2 galaxies are stronger 25 and 60 $\mu$m emitters, have a
larger far-IR to optical excess \( \frac{L_{IR}}{L_B} \) and steeper (redder) 
12-25 $\mu$m and 25-60 $\mu$m spectra, compared to Seyfert 1s.
These differences suggest larger amounts of dust or/and larger 
contribution of star formation in Seyfert 2s, both indicating {\em intrinsic} 
differences between the two Seyfert classes. 

(ii) In Seyfert 2s, the mid- and far-IR emission correlate with {\em total} 
optical luminosities and host galaxy sizes, this indicating that the bulk of IR 
emission in these objects is dominated by warm dust in the host galaxy disks. 
We find an anti-correlation between 25 $\mu$m emission and nuclear 
ionization in Seyfert 2s, which is contradicting the simple orientation 
unification scheme where the 25 $\mu$m emission is dominated by the AGN. At high 
IR powers, Seyfert 2s tend to be optically fainter than Seyfert 1s, for the same 
far-IR brightness, this being consistent with larger dust content in Seyfert 2s, 
especially at nuclear scales. Various correlations between IR colours and optical
luminosities suggest that that the 25-60 $\mu$m emission in these
objects is primarily originating in their disks. Shortward of 25 $\mu$m their 
IR spectra are dominated by nuclear scales.

(iii) Seyfert 1s span a very narrow range in $L_{25}$ over a large range in 
optical luminosities, with no obvious correlation between them at any spatial 
scales or wavelengths. The narrow 25 $\mu$m emission range in these objects might
be indicating the emitting dust properties. The far-IR emission scales with 
{\em total} galaxian light and colours and with host size, this indicating that 
the Seyfert 1 far-IR emission is disk-dominated, as in the case of Seyfert 2s. 
There is no obvious correlation between the mid-to-far IR spectral shapes and 
the optical properties of Seyfert 1s, at any spatial scales.

(iv) Seyfert 2s are IR-louder, both at 25 and 60 $\mu$m, compared to Seyfert 1s, 
especially at nuclear scales. IR-loudness correlates with brighter {\em total} 
optical luminosities and bluer {\em total} colours in Seyfert 2s, but with 
fainter optical luminosities and redder colours, at all spatial scales, in 
Seyfert 1s. In the latter objects, total IR-loudness also anti-correlates with 
host size. These results indicate that in Seyfert 1s IR-loudness is a 
measure of dust extinction and host luminosity, while in Seyfert 2s it is 
primarily indicative of IR excess emission at large scales and affected by dust 
extinction at nuclear scales.

(v) The Warm sample as a whole shows stronger 25 and weaker 60 $\mu$m 
emission, compared to the Cold sample. Cold galaxies are IR-louder at 
60 $\mu$m, especially on nuclear scales. These results indicate both larger 60
$\mu$m emission and nuclear dust absorption in IR-Cold galaxies. 
IR luminosities of Cold galaxies correlate with galaxy size, which indicates 
that their IR emission is dominated by warm dust in their disks. The closely 
interacting systems among them, tend to have larger IR excess.
We find no significant correlations between the mid- or far-IR and optical 
luminosities
of Cold galaxies, at all spatial scales. We interpret this as a result
of dust extinction affecting their optical properties at all spatial scales.
The 12-100 $\mu$m colour index is as a good separator between Warm and Cold
samples, as is the 25-60 $\mu$m colour index and scales with \( \frac{L_{IR}}{L_B} \)
in Cold galaxies. In Cold galaxies, the spectral shape at 12 $\mu$m is correlated with nuclear
optical properties, while longward of 60 $\mu$m with their total optical
properties. On the other hand we find no significant correlations involving
the 25-60 $\mu$m colour index in these objects.

(vi) In summary, we find a transition from fainter to brighter nuclear magnitudes,
redder to bluer nuclear optical colours, redder to bluer 12-100 colours, larger 
to smaller \( \frac{L_{IR}}{L_B} \) and decreasing nuclear IR-loudness, as we
progress from the Cold, to the Warm Seyfert 2 and to the Warm Seyfert 1 galaxies.

\section{Concluding Remarks}

In this paper we presented the results of aperture photometry for 
our IR-Warm and Cold samples and discussed their optical properties (luminosities
and colours, at various spatial scales and host sizes) in correlation with their 
IR properties (luminosities and colours). Our main conclusions are summarized in
Sections 2.3 and 3.6. We find a transition in the observed optical and IR
properties from Cold to Warm Seyfert 2 and Seyfert 1 galaxies,
with a partial overlap between the first two. 

Beginning to answer the two main questions that we formulated in Paper I, 
testing the universality of the orientation model and investigating the
origin of mid-IR excess in Seyfert galaxies, our 
data so far indicates that: (i) Although some of the observed differences between
the Warm Seyfert 1 and 2 samples (fainter and redder nuclei for the latter) could
be accounted for by a preferential obscuration of the Seyfert 2 nuclei, other
properties (larger and brighter Seyfert 2 disks) cannot, but rather indicate
intrinsic differences between the two Seyfert types. In fact, the correlations between
optical and IR properties indicate larger dust content and probably disk star 
formation in Seyfert 2s. (ii) The 25 $\mu$m emission in our Warm Seyferts is
mainly due to the AGN thermal component, being further enhanced in Seyfert 2s by
star formation. Their far-IR (60-100 $\mu$m) emission, on the other hand, 
originates in their host disks. In Cold galaxies, the bulk or IR emission at all
wavelengths is dominated by warm dust in their disks, probably heated by strong
star formation (at least in strongly interacting systems).

\acknowledgments
I am grateful to my thesis advisors George Miley and Walter Jaffe for providing
me with stimulation and support throughout the completion of this project.
This research has made use of the NASA/IPAC Extragalactic Database (NED) 
which is operated by the Jet Propulsion Laboratory, California Institute of 
Technology, under contract with the National Aeronautics and Space 
Administration. Part of this work was completed while the author held a 
National Research Council - NASA GSFC Research Associateship.

%
%

\clearpage
\begin{deluxetable}{lrrrrrr}
\tablecaption{Aperture Photometry of Cold sources. \label{tab2}}
\tablewidth{0pt}
\tablehead{
\colhead{Identification} & \colhead{$R$} & \colhead{$(B-V)$} & \colhead{$(B-R)$} & \colhead{$(V-R)$} & \colhead{$(V-I)$} & \colhead{$D_{25}$} \\
\colhead{} & \colhead{(mag)} & \colhead{(mag)} & \colhead{(mag)} & \colhead{(mag)} & \colhead{(mag)} & \colhead{(kpc)}
}
\startdata
     IRAS 07514+5327 (I231) & 12.92$\pm$0.07 & 0.39$\pm$0.10 & 1.16$\pm$0.10 & 0.77$\pm$0.09 & \nodata & 36.0 \nl
                (north nucleus)    & 15.13$\pm$0.07 & 0.61$\pm$0.10 & 1.15$\pm$0.10 & 0.54$\pm$0.09 & \nodata & 2 \nl
                (south nucleus)    & 14.97$\pm$0.07 & 0.66$\pm$0.10 & 1.20$\pm$0.10 & 0.54$\pm$0.09 & \nodata & 2 \nl
     IRAS 06506+5025 (I211) & 13.54$\pm$0.08 & 0.70$\pm$0.10 & 1.49$\pm$0.10 & 0.79$\pm$0.09 & \nodata & 15.3 \nl
                                   & 14.40$\pm$0.08 & 0.54$\pm$0.10 & 1.50$\pm$0.10 & 0.95$\pm$0.09 & \nodata & 2 \nl
     IRAS 02439-7455 (Fairall11) & 13.98$\pm$0.01 & \nodata & 0.53$\pm$0.10 & \nodata & \nodata & 24.6 \nl 
                                   & 14.59$\pm$0.01 & \nodata & 0.46$\pm$0.10 & \nodata & \nodata & 2 \nl
       IRAS 04015-1118 (NGC1509) & 12.90$\pm$0.01 & \nodata & 0.87$\pm$0.10 & \nodata & \nodata & 28.6 \nl  
                                   & 13.80$\pm$0.01 & \nodata & 0.75$\pm$0.10 & \nodata & \nodata & 2 \nl
       IRAS 05217-4245 (ESO252-G022) & 13.27$\pm$0.01 & \nodata & 1.64$\pm$0.10 & \nodata & \nodata & 44.4 \nl
                                   & 15.36$\pm$0.01 & \nodata & 1.96$\pm$0.10 & \nodata & \nodata & 2 \nl
       IRAS 04265-4801 (ESO202-G023) & 12.20$\pm$0.01 & \nodata & 1.34$\pm$0.10 & \nodata & \nodata & 31.8 \nl
                                   & 14.33$\pm$0.01 & \nodata & 2.05$\pm$0.10 & \nodata & \nodata & 1 \nl 
       IRAS 10475-1429 (MGC02-28-009)W & 14.14$\pm$0.01 & \nodata& 1.35$\pm$0.10 & \nodata & \nodata & 29.5 \nl 
                                   & 15.96$\pm$0.01 & \nodata & 1.52$\pm$0.10 & \nodata & \nodata & 2 \nl
       IRAS06+1018 (CGCG063-051)N & 15.56$\pm$0.01 & \nodata & 0.77$\pm$0.10 & \nodata & \nodata & 34.0 \nl
                                   & 17.54$\pm$0.01 & \nodata & 1.12$\pm$0.10 & \nodata & \nodata & 2 \nl
       IRAS 09406+1018 (CGCG063-051)S & 16.74$\pm$0.02 & \nodata & 1.56$\pm$0.10 & \nodata & \nodata & 5.5 \nl
                                   & 17.19$\pm$0.01 & \nodata & 1.66$\pm$0.10       & \nodata & \nodata & 2 \nl
       IRAS 04530-3850 (ESO304-IG029)N & 14.60$\pm$0.01 & \nodata & \nodata & 0.58$\pm$0.01 & \nodata & 30.3\tablenotemark{*} \nl 
                                   &  17.34$\pm$0.01 & \nodata & \nodata & 0.62$\pm$0.01 & \nodata & 2 \nl
       IRAS 04530-3850 (ESO304-IG029)S & 14.61$\pm$0.01 & \nodata & \nodata & 0.43$\pm$0.02 & \nodata & 34.1\tablenotemark{*} \nl
                                   & 15.76$\pm$0.01 & \nodata & \nodata & 0.49$\pm$0.01 & \nodata & 2 \nl
       IRAS 03531-4507 (ESO249-IG028) & 15.02$\pm$0.01 & \nodata & 0.85$\pm$0.10 & \nodata & \nodata & 31.7 \nl
                                   & 16.44$\pm$0.01 & \nodata & 0.85$\pm$0.10 & \nodata & \nodata & 2 \nl
       IRAS 04304-5323 (ESO157-IGA040)N & 15.09$\pm$0.01 & \nodata & \nodata & 0.56$\pm$0.01 & \nodata & 35.8\tablenotemark{*} \nl 
                                   & 17.27$\pm$0.01 & \nodata & \nodata & 0.65$\pm$0.01 & \nodata & 2 \nl
       IRAS 04454-4838 (ESO203-IG001)N & 15.89$\pm$0.01 & \nodata & \nodata & 0.48$\pm$0.01 & \nodata & 21.5\tablenotemark{*} \nl 
                                   & 17.19$\pm$0.01 & \nodata & \nodata & 0.47$\pm$0.01 & \nodata & 2 \nl
       IRAS 23179-6929 (ESO077-IG014) & 14.62$\pm$0.01 & \nodata & \nodata & 0.66$\pm$0.01 & \nodata & 25.9\tablenotemark{*} \nl
                                   & 15.98$\pm$0.01 & \nodata & \nodata & 0.80$\pm$0.01 & \nodata & 2 \nl
       IRAS 05207-2727 (ESO487-IG006) & 14.42$\pm$0.01 & \nodata & 1.50$\pm$0.10 & \nodata & \nodata & 33.3 \nl  
                                   & 16.32$\pm$0.10 & \nodata & 1.86$\pm$0.10 & \nodata & \nodata & 2 \nl
\enddata
\tablenotetext{*} {Diameter of the 25th mag arcsec$^{-2}$ isophote measured on the $V$ image}
\tablecomments{Total and Nuclear Magnitudes and Colours. Sources in order of decreasing $\alpha_{(25,60)}$}

\end{deluxetable}

\clearpage

\begin{deluxetable}{lrrrrrrrr}
\tablecolumns{9}
\tablewidth{0pc}
\tablecaption{Median and Mean Quantities for all Samples. \label{tab3}}
\tablehead{
\colhead{}    &  \multicolumn{4}{c}{Median} &  \multicolumn{4}{c}{Mean} \\
\cline{2-5} \cline{6-9} \\
\colhead{Quantity} & \colhead{Seyfert 1} & \colhead{Seyfert 2} & \colhead{Warm} & \colhead{Cold} & \colhead{Seyfert 1} & \colhead{Seyfert 2} & \colhead{Warm} & 
\colhead{Cold} \\
}
\startdata
$B_{nuc}$\tablenotemark{1} & 15.85 & 16.98 & 16.81 & 17.29 & 15.92 & 16.83 & 16.49 & 16.90 \nl
$V_{nuc}$ & 15.27 & 16.27 & 15.69 & 16.78 & 15.27 & 16.15 & 15.82 & 16.79 \nl
$R_{nuc}$ & 15.16 & 15.83 & 15.24 & 15.98 & 14.88 & 15.59 & 15.33 & 15.91 \nl
$I_{nuc}$ & 14.48 & 15.33 & 14.71 & \nodata & 14.37 & 14.96 & 14.66 & \nodata \nl
$B_{25}$\tablenotemark{2} & 14.97 & 15.54 & 15.42 & 15.03 & 15.25 & 15.19 & 15.21 & 15.25 \nl
$V_{25}$ & 14.60 & 15.03 & 14.68 & 15.20 & 14.59 & 14.65 & 14.63 & 15.08 \nl
$R_{25}$ & 14.15 & 14.33 & 14.32 & 14.60 & 14.17 & 14.17 & 14.17 & 14.34 \nl
$I_{25}$ & 13.66 & 13.78 & 13.66 & \nodata & 13.63 & 13.32 & 13.47 & \nodata \nl
$B_{disk}$\tablenotemark{3} & 16.45 & 15.81 & 16.16 & 15.68 & 16.25 & 15.53 & 15.80 & 15.67 \nl
$V_{disk}$ & 15.69 & 15.17 & 15.39 & 15.47 & 15.56 & 15.03 & 15.23 & 15.38 \nl
$R_{disk}$ & 15.20 & 14.83 & 15.18 & 14.89 & 15.09 & 14.59 & 14.77 & 14.74 \nl
$I_{disk}$ & 14.68 & 13.89 & 14.23 & \nodata & 14.53 & 13.65 & 14.09 & \nodata \nl
$(B-V)_{nuc}$ & 0.29 & 0.87 & 0.81 & 0.63 & 0.42 & 0.88 & 0.70 & 0.59 \nl
$(B-R)_{nuc}$ & 0.77 & 1.49 & 1.37 & 1.50 & 0.89 & 1.46 & 1.24 & 1.36 \nl
$(V-R)_{nuc}$ & 0.49 & 0.62 & 0.57 & 0.62 & 0.49 & 0.60 & 0.56 & 0.65 \nl
$(V-I)_{nuc}$ & 0.84 & 1.21 & 1.04 & \nodata & 0.79 & 1.16 & 0.97 & \nodata \nl
$(B-V)_{25}$ & 0.41 & 0.63 & 0.49 & 0.70 & 0.46 & 0.59 & 0.54 & 0.55 \nl
$(B-R)_{25}$ & 0.90 & 1.29 & 1.27 & 1.34 & 0.94 & 1.25 & 1.13 & 1.19 \nl
$(V-R)_{25}$ & 0.49 & 0.59 & 0.57 & 0.60 & 0.49 & 0.65 & 0.59 & 0.61 \nl
$(V-I)_{25}$ & 0.94 & 1.07 & 1.02 & \nodata & 0.90 & 1.09 & 0.99 & \nodata \nl
$(B-V)_{disk}$ & 0.51 & 0.49 & 0.49 & 0.82 & 0.51 & 0.49 & 0.50 & 0.58 \nl
$(B-R)_{disk}$ & 0.99 & 1.24 & 1.11 & 1.26 & 1.00 & 1.16 & 1.10 & 1.16 \nl 
$(V-R)_{disk}$ & 0.50 & 0.58 & 0.56 & 0.59 & 0.49 & 0.66 & 0.59 & 0.59 \nl
$(V-I)_{disk}$ & 1.04 & 1.07 & 1.04 & \nodata & 1.07 & 1.08 & 1.07 & \nodata \nl
$D_{25}[B+V]$ & 26.8 & 31.0 & 30.7 & 31.7 & 26.15 & 32.06 & 29.80 & 30.45 \nl
$D_{25}[B]$ & 28.9 & 35.8 & 32.2 & 31.8 & 27.01 & 33.34 & 30.81 & 30.92 \nl
\enddata
\tablenotetext{1} {Within 2 kpc from the nucleus}
\tablenotetext{2} {Within elliptical apertures corresponding to $\mu_{B}$=25 mag arcsec}
\tablenotetext{3} {Within the annulus defined between the above apertures}
\end{deluxetable}

\begin{deluxetable}{lrrrr}
\tablewidth{0pt}
\tablecaption{Mean IR-loudness indices. \label{tab4}}
\tablehead{
\colhead{Quantity} & \colhead{Seyfert 1} & \colhead{Seyfert 2} & \colhead{Warm} & \colhead{Cold} \\
}
\startdata
		$\alpha_{(V,25)}$[25] & -1.11 & -1.23 & -1.18 & -1.12 \nl
		$\alpha_{(V,25)}$[nuc] & -1.27 & -1.59 & -1.47 & -1.53 \nl
		$\alpha_{(V,60)}$[25] & -1.02 & -1.13 & -1.09 & -1.40 \nl
		$\alpha_{(V,60)}$[nuc] & -1.16 & -1.43 & -1.32 & -1.74 \nl
\enddata
\tablecomments{The subscripts [nuc] and [25] refer to $V$ magnitudes measured within the central 2 kpc and $\mu_{B}$= 25 mag arcsec$^{-2}$ respectively (see text).}
\end{deluxetable}

%
%

%
%

\clearpage

\clearpage

\clearpage

\clearpage

\clearpage

\clearpage

\clearpage

\end{document}